\documentstyle[preprint,aps]{revtex}

\begin{document}
\draft
\title{Tunneling Between Two-Dimensional Electron Gases in a
Strong Magnetic Field}
\author{K.~M. Brown, N. Turner, J.~T. Nicholls, E.~H. Linfield,
M. Pepper, D.~A. Ritchie, and G.~A.~C. Jones}
\address{Cavendish Laboratory, Madingley Road, Cambridge CB3 0HE, England}

\date{\today}
\maketitle

\begin{abstract}
We have measured the tunneling between two two-dimensional
electron gases at high magnetic fields $B$,
when the carrier densities of the two electron layers are matched.
For filling factors $\nu<1$, there is a gap in the current-voltage
characteristics centered about $V=0$,
followed by a tunneling peak at $\sim 6$~mV.
Both features have been observed
before and have been attributed
to electron-electron interactions within a layer.
We have measured high field tunneling peak positions and fitted gap parameters
that are proportional to $B$,
and independent of the carrier densities of the two layers.
This suggests a different origin for the gap
to that proposed by current theories,
which predict a $\sqrt{B}$ dependence.
\end{abstract}


\narrowtext

Following recent experimental work\cite{ash90,demm91,eis92a}
there has been much theoretical interest in the tunneling of
electrons out of a two-dimensional electron gas (2DEG)
in a strong magnetic field.\cite{yang93,hats93,song93a,johan93,efros93,varma94}
Following on from experiments of Demmerle {\it et al.\/}\cite{demm91},
Eisenstein {\it et al.\/}\cite{eis92a,eis94a}
measured the tunneling from one 2DEG to a similar parallel 2DEG,
the two being separated by a 175~\AA\ barrier.
In zero magnetic field there is resonant tunneling between
the two 2DEGs when their carrier densities are equal.
In a strong magnetic field however,
the current-voltage ($I$-$V$) characteristics between the
two layers exhibit a gap when the filling
factor $\nu$ is less than unity.
This suppression of tunneling has been interpreted as evidence for
electron-electron interactions within a 2DEG,
and the resulting gap was labeled a ``Coulomb gap''.
It was also suggested\cite{eis92a} that,
at high magnetic fields, electron-electron interactions are
responsible for lowering the effective mass of electrons
in the lowest Landau level (LL).

In this paper we present tunneling results
obtained on similar double 2DEG samples based on GaAs.
We observe a gap in the $I$-$V$ characteristics,
but we shall show that the gap and the associated tunneling peak
do not follow the $\sqrt{B}$ behavior
predicted by current theories.
Instead we observe a high field gap and a tunneling
peak that are both linear in $B$ and
independent of filling factor.

Double quantum wells and buried patterned back gates
were fabricated by {\it in situ\/} ion beam
lithography and molecular beam epitaxy regrowth.
These techniques allow patterned back gates
to be grown into the wafer structure,
the details of which have
been published elsewhere.\cite{lin93,brown94a,brown94b}
Subsequent optical lithography was used to define a Hall bar mesa,
and to deposit Au Schottky gates aligned with the back gates.
The device was put into the tunneling configuration
(as shown in the left hand inset of Fig.~\ref{fig1})
by applying negative voltages to side front and back gates (not shown).
The two electron gases then overlap in a 100~$\mu$m $\times$ 150~$\mu$m area.
The carrier densities of the top ($n_1$) and bottom ($n_2$) 2DEGs
in the tunneling area were controlled by voltages applied to the top
($V_{\text{g1}}$) and bottom ($V_{\text{g2}}$) gates, respectively.

We present results obtained from one particular wafer (C751),
which consists of two modulation-doped 180~\AA\ GaAs
quantum wells separated by a 125~\AA\
Al$_{0.33}$Ga$_{0.67}$As barrier.
The upper 2DEG has an as-grown carrier density of
$n_1=3.1\times 10^{11}$~cm$^{-2}$
and a mobility of $8\times 10^5$~cm$^2$/Vs,
the corresponding quantities for the lower
2DEG are $n_2=1.8\times 10^{11}$~cm$^{-2}$
and $2\times 10^5$~cm$^2$/Vs.
The surface gates were evaporated 800~\AA\ above the top 2DEG,
whereas the buried gates were fabricated
3500~\AA\ below the bottom 2DEG.

Figure~\ref{fig1} shows the $I$-$V$
characteristics of sample C751
measured at 35~mK from 13 to 16~T,
when the carrier densities in the tunneling region were set to
$n_1=n_2=1.9 \times 10^{11}$~cm$^{-2}$.
In contrast to the resonant tunneling that occurs at zero magnetic field,
there is a strong suppression of the tunneling
around $V=0$ at high magnetic fields.
A suppression of the zero bias tunneling
(though not enough to form a gap)
has also been observed at filling factors as high as $\nu=8$,
but we only show data for $\nu<1$.
Our samples have a lower mobility than those
used in previous studies,\cite{eis92a}
but the features that we observe are the same,
though not so well defined.
One advantage of the lower mobility samples is that
the high field measurements ($\nu<1$) were not affected
by fractional quantum Hall states.

The data in Fig.~\ref{fig1} show a
gap centered at $V=0$ which gives way to a
tunneling peak (labeled $E_1$) at $\sim$5-7~mV,
and a higher peak at $\sim 30$~mV (labeled $E_2$).
The additional splitting of the $E_2$ peak around 13~T is due
to a peak that moves to lower voltage as the magnetic field is increased.
This extra peak is due to inter LL transitions
originating from the first excited 2D subband of the quantum wells;
we shall concentrate on results only within the lowest 2D subband.
In the right hand inset of Fig.~\ref{fig1}
we have plotted the $E_1$ and $E_2$ peak positions between 9-16~T.
The two sets of data points have been fitted with straight lines
forced through the origin, with slopes of
$E_1=(0.30 \pm 0.04) \hbar\omega_c$
and $E_2=(1.3 \pm 0.1) \hbar \omega_c$,
where the cyclotron energy is $\hbar\omega_c=1.67$~meV/T for GaAs.

Additional information about the gap can be obtained
from the functional form of the $I$-$V$ characteristics.
Song He {\it et al.\/}\cite{song93a}
predict that the tunneling
current at $\nu=1/2$ will have the form $I=I_0 \exp (-\Delta/V)$,
where the gap parameter $\Delta=2\pi e^2/\epsilon l_B$,
and $l_B = \sqrt{\hbar/eB}$ is the magnetic length.
In another description of this system,
Ashoori {\it et al.\/}\cite{ash93} assumed a linear
variation of the density of states about $E_{\text{F}}$,
and argue that the tunneling current
should have the form $I \sim V^3$.
Johansson {\it et al.\/}\cite{johan93}
predict a similar form, but of a higher power.

Experimentally we cannot fit the low voltage
tunneling current with any polynomial.
Instead, we find that the function $I=I_0 \exp (-\Delta/V)$ fits
the tunneling data best in the high bias regime,
a surprising result given that the theory\cite{song93a}
should only be applicable in the low energy limit, $V \rightarrow 0$.
The inset to Fig.~\ref{fig2} shows
an $I$-$V$ characteristic at 16~T and $\nu=1/2$
plotted as $\log(\mid I \mid)$ {\it vs.} 1/$V$;
equally good fits were obtained over a wide
range of filling factors $0.2<\nu<2$.
The values of $\Delta$ extracted from the $I$-$V$
sweeps in Fig.~\ref{fig1} are plotted along with
$E_1$ and $E_2$ in the figure inset.
The gap parameter $\Delta$ is best fit by
the function $\Delta=0.44 \hbar \omega_c$.

Theoretical descriptions of the gap and the $E_1$ tunneling peak
can be subdivided into those that ignore
interlayer correlations,\cite{yang93,hats93,song93a,johan93}
and those that incorporate interlayer interactions.\cite{efros93,varma94}
Quantum mechanical calculations\cite{hats93,song93a} of the former type predict
that,
for the compressible state at $\nu=1/2$,
the tunneling peak occurs at the energy $0.4e^2/\epsilon l_B \sim \sqrt{B}$.
The Coulomb energy $0.4e^2/\epsilon l_B$
(shown as a dashed line in the Fig.~\ref{fig1} inset)
is comparable to the energy of the $E_1$ peak position.
Our data shows however that the $E_1$ peak
position is better fit by a linear,
rather than a square root, function of magnetic field $B$.

To test theoretical
predictions\cite{song93a} at $\nu=1/2$,
we performed a series of experiments at
magnetic fields and carrier densities ($n_1=n_2$),
such that the lowest LL in each layer was always half-filled.
At filling factor $\nu=1/2$, the two 2DEGs in the tunneling
region are Fermi liquids,\cite{halp93}
and complications due to the current flowing
along the edges of the tunneling region are avoided.
We have investigate this filling factor over a wide range of magnetic fields,
and Fig.~\ref{fig2} shows the value of the gap parameter
$\Delta$ measured at $\nu=1/2$ from 3 to 16~T.
The up and down triangles show the fitted
values of $\Delta$ in the two bias directions.
Not only is the magnitude of $\Delta$
much smaller than theoretically predicted
($2 \pi e^2/\epsilon l_B=85$~meV at 10~T),
but the average value of $\Delta$
is best fit by the function $\Delta=(0.44 \pm 0.02) \hbar \omega_c$,
which is proportional to $B$ rather than $\sqrt{B}$.
Similar results were obtained at filling factors $\nu=0.4$ and 0.6,
showing that the choice of $\nu=1/2$ is not special.

Johansson {\it et al.\/},\cite{johan93} modeled
the electron liquids in each of the
two layers as Wigner crystals.
The energy of the tunneling peak is predicted to be equal to the
intralayer Coulomb energy $E_C=e^2/\epsilon a$,
where $a$ is the lattice parameter of the Wigner crystal
determined by the 2D electron density $n$.
$E_C$ is comparable to $E_1$,
but the model predicts that the energy of the peak maximum
depends only on the carrier densities $n_1$ and $n_2$.
Figure \ref{fig3} shows the $I$-$V$
characteristics measured
when the magnetic field was fixed at $B=16$~T,
while the carrier densities $n=n_1=n_2$ were varied
from 1 to $3 \times 10^{11}$~cm$^{-2}$.
The $E_1$ and $E_2$ peak positions remained constant as
the filling factor was changed by a factor of three.
A similar independence of the gap with electron density
has also been observed by Ashoori {\it et al.}\cite{ash90,ash93}
The $E_1$ peak occurs at 0.3$\hbar \omega_c$ and the
$E_2$ peak remains at approximately 1.3$\hbar \omega_c$,
the same values that were obtained when the
carrier densities were fixed and the magnetic
field was varied (see Fig.~\ref{fig1}).
The fitted gap parameter $\Delta$
remained constant 0.44$\hbar \omega_c$.
We find no evidence that the $E_1$ peak position
scales with carrier density as $E_1 \sim \sqrt{n}$,
as predicted by Johansson {\it et al.}\cite{johan93}
In a model that also incorporates interlayer electron correlations,
Varma {\it et al.\/}\cite{varma94} have also predicted that the gap
will be a strong function of the interelectron distance within a plane;
the results in Fig.~\ref{fig3} show this not to be the case.

In previous experiments\cite{demm91} at 4.2~K,
the position of the maxima in $dI/dV$ were used to
identify inter LL transitions that occur
when the bias energy $eV =\pm \hbar \omega_c, \pm 2\hbar \omega_c$.
There is a maximum when there is tunneling from the $N=0$ LL in one 2DEG,
to the $N=1,2$ LL in the other 2DEG.
There was no indication
of a suppression of tunneling at zero bias in these experiments,
possibly due to the high measuring temperature.
In experiments\cite{eis92a} that extended this
earlier work to lower temperatures
and higher magnetic fields, a gap was measured.
The $E_2$ tunneling peak was interpreted as an inter LL transition,
occurring at $1.3\hbar\omega_c$ rather than at the
usual cyclotron energy $\hbar\omega_c$.
It was suggested that the increased cyclotron energy was
due to a decrease of the effective mass resulting from
intralayer Coulomb interactions.\cite{smith92}

We have investigated other samples with different
mobilities and AlGaAs barrier widths,
and have obtained similar tunneling results to
those shown in Figs.~\ref{fig1}-\ref{fig3}.
For barrier widths where tunneling is important (100-200~\AA),
we observe linear magnetic field dependent
behavior for $E_1$, $E_2$, and $\Delta$.
The energies $E_1$ and $E_2$ have roughly the same values as in C751,
but the fitted values of $\Delta$ are larger in more disordered samples.
In all cases, by whatever criterion we choose
to describe the $E_1$ tunneling peak
(for example, by an onset voltage or a peak position),
we always obtain a linear magnetic field
dependence for the quantity chosen.
These results suggest that the $E_1$ tunneling peak,
like the $E_2$ peak, could be related to the Landau
level structure within each 2DEG.
There is however no obvious LL structure that would give
rise to a peak at 0.3$\hbar \omega_c$.
If spin-splitting is taken into account then a spin-flip transition of
$2 g \mu_B SB =0.3 \hbar \omega_c$ would suggest a $g$-value of 4.5;
such a large $g$-value that is independent of $\nu$ is not expected in GaAs.
Moreover, spin splitting cannot account for
the suppression of tunneling about zero bias.

Theoretical descriptions of electron-electron
interactions in a strong magnetic field,
for example, the fractional quantum Hall effect,
work within the lowest Landau level approximation
$\hbar \omega_c \gg e^2/\epsilon l_B$.
The energy gap for a fractional state (for example, $\nu=1/3$) is expected
to be proportional to the energy $e^2/\epsilon l_B$.
Various calculations have included the
effect of higher LLs and have predicted a departure
from $\sqrt{B}$ towards linear behavior for the fractional gap.
We could be observing an analogous effect
for the tunneling gap between two 2DEGs,
though as we suggest in the next paragraphs
the explanation could be more mundane.

We have been able to follow{\cite{brown94d}
the $E_1 \approx 0.3 \hbar \omega_c$ tunneling
peak down to low magnetic fields ($B=1$~T, $\nu=8$).
It would be suprising if we could detect intralayer electron
correlations at such high filling factors,
and a classical description of the system may be more appropriate.
In a classical theory put forward by Efros {\it et al.},\cite{efros93}
the Coulomb gap arises due to correlations
both between and within the electron layers.
The classical theory predicts\cite{efros93} that,
for fixed carrier densities $n_1=n_2$,
the gap $\Delta$ is linear in $\nu$ for $0.3 < \nu < 0.9$;
our data in Fig.~\ref{fig1}
show that $\Delta$ is linear in $B$.
The classical theory also predicts that, for a given filling factor,
the gap $\Delta$ scales with $\sqrt{B}$;
the data at $\nu=1/2$ in Fig.~\ref{fig2}
show that this is not the case.

The available theories do not describe our data,
at constant carrier densities they predict a gap that is either
proportional to $\sqrt{B}$ or to $\nu$.
Our data shows that the gap only depends on $B$.
In further experiments\cite{brown94e} we have measured the
$I$-$V$ characteristics when the lower 2DEG was fixed at
$n_2=3.1 \times 10^{11}$~cm$^{-2}$ at $B=16$~T.
The $E_1$ and $E_2$ tunneling peaks and the fitted gap $\Delta$
remained constant as $n_1$ was varied from
$3.1 \times 10^{11}$~cm$^{-2}$ to $1.1 \times 10^{11}$~cm$^{-2}$.
This result, as well as those presented in Figs.~\ref{fig1}-\ref{fig3},
strongly suggest a classical origin for the gap.
In this is the case then the $E_2$ peak can be explained as being the usual
inter LL transition at $\hbar \omega_c$ which has been displaced by
0.3$\hbar \omega_c$ to $1.3\hbar \omega_c$.

In conclusion, we have observed a suppression of
tunneling between between two partially filled LLs.
With our samples we have been able to
extend tunneling measurements over a wide range of
carrier densities and filling factors.
Providing that the sample does not freeze out at high magnetic field,
our results show that the gap solely depends on the magnetic field $B$,
and is independent of mobility and carrier densities of the two layers.
We show that the $E_1$ peak energy is not
proportional to the electron-electron interaction
energies $e^2/\epsilon l_B$ or $e^2/\epsilon a$.

We wish to thank the Engineering and Physical Sciences
Research Council (UK) for supporting this work.
JTN acknowledges support from the Isaac Newton Trust,
and DAR acknowledges support from Toshiba Cambridge Research Centre.

\begin{figure}
\caption{Main figure: low temperature $I$-$V$ characteristics
when $n_1=n_2=1.9 \times 10^{11}$~cm$^{-2}$.
Sweeps were measured at 0.2~T intervals,
and those at 15.4~T and 14.8~T are missing.
Left hand inset: the device in the tunneling configuration.
Right hand inset: fitted gap $\Delta$ and
peak positions $E_1$ and $E_2$ as a function of magnetic field $B$.
The solid lines are the fits $E_1=0.30\hbar \omega_c$,
$E_2=1.3\hbar \omega_c$, and $\Delta=0.42\hbar \omega_c$.
The dashed line shows the Coulomb energy $0.4 e^2/\epsilon l_B$.}
\label{fig1}
\end{figure}

\begin{figure}
\caption{The fitted gap parameter $\Delta$ measured at
$\nu=1/2$ as a function of magnetic field.
The solid line is a least-squares fit to the average value of
$\Delta$ in the two bias directions.
Inset: $\log(\mid I \mid)$ {\it vs.\/} 1/$V$ for
$n_1=n_2=1.94 \times 10^{11}$~cm$^{-2}$ at $B=16$~T,
showing the range of fit to the function $I=I_0 \exp(-\Delta/V)$.}
\label{fig2}
\end{figure}

\begin{figure}
\caption{$I$-$V$ characteristics measured at $B=16$~T,
as the filling factor was varied from 0.26 to 0.77.}
\label{fig3}
\end{figure}

\end{document}